\documentclass[conference]{IEEEtran}
\usepackage{cite}
\usepackage{amsmath,amssymb,amsfonts}
\usepackage{algorithmic}
\usepackage{graphicx}
\usepackage{textcomp}
\usepackage{xcolor}
\usepackage{amsmath} 
\usepackage{tikz}
\usetikzlibrary{positioning}
\usepackage[margin=1in]{geometry}
\usepackage{xcolor}
\def\BibTeX{{\rm B\kern-.05em{\sc i\kern-.025em b}\kern-.08em
    T\kern-.1667em\lower.7ex\hbox{E}\kern-.125emX}}
\begin{document}

\title{FlashGuard: Novel Method in Evaluating Differential Characteristics of Visual Stimuli for Deterring Seizure Triggers in Photosensitive Epilepsy\\
}

\author{\IEEEauthorblockN{1\textsuperscript{st} Ishan Pendyala}
\IEEEauthorblockA{
pendyala.ishan@gmail.com
\\
Friendswood High School \\
0009-0008-0086-8287}

}

\maketitle

\begin{abstract}
In the virtual realm, individuals with photosensitive epilepsy (PSE) encounter challenges when using devices, resulting in exposure to unpredictable seizure-causing visual stimuli. The current norm for preventing epileptic flashes in media is to detect asynchronously when a flash will occur in a video, then notifying the user. However, there is a lack of a real-time and computationally efficient solution for dealing with this issue. To address this issue and enhance accessibility for photosensitive viewers, FlashGuard, a novel approach, was devised to assess the rate of change of colors in frames across the user's screen and appropriately mitigate stimuli, based on perceptually aligned color space analysis in the CIELAB color space. The detection system is built on analyzing differences in color, and the mitigation system works by reducing luminance and smoothing color transitions. This study provides novel insight into how intrinsic color properties contribute to perceptual differences in flashing for PSE individuals, calling for the adoption of broadened WCAG guidelines to better account for risk. These insights and implementations pave the way for stronger protections for individuals with PSE from dangerous triggers in digital media, both in policy and in software.

\end{abstract}


\section{Introduction}
The virtual realm poses a problem to the large majority of individuals suffering from photosensitive epilepsy (PSE), based primarily on the large amounts of unmonitored content that inhabit social media sites and video-displaying sites. A further, unresolved problem stems from the advent of the most consumed forms of content being displayed in real-time, such as on "rapid-consumption" apps (like TikTok or Instagram Reels), where potentially dangerous visual stimuli are shown to the PSE-individual, either accidentally or purposefully \cite{b1, b4}.

Another large source of PSE-triggering stimuli comes in other forms of digital media in everyday computer usage, which accounts for a large proportion of PSE-related seizures in adolescents. For example, although 60\% of patients with PSE are female, males account for a higher proportion of recorded seizures, hypothesized to be due to greater exposure to video games or media containing flashing lights with rapid flash without warning \cite{b1, b2}.

Although published standards for safety thresholds in content with flashing exist-most notably, the Web Content Accessibility Guidelines (WCAG) 2.0 \cite{b3}, these recommendations are not enforced across most video platforms, and therefore becomes the responsibility of the user to mitigate risks, necessitating a "consumer-driven protection" mechanism in which those watching the content must protect themselves \cite{b4}. Because of this, PSE patients continue to experience risk from unregulated content, often in everyday online interactions. Additionally, as this study argues, the WCAG Success Criterion in the “Three Flashes or Below Threshold” (Success Criterion 2.3.1) \cite{b3}, last updated in 2008, is insufficient in wholly mitigating seizure risk, given its specific focus on only luminance flashes and saturated-red flashes. 

The Photosensitive Epilepsy Analysis Tool (PEAT) has been used to determine seizure risk by other researchers, as it analyzes media in accordance with WCAG Success Criterion 2.3.1 \cite{b5}. While PEAT is largely unusable by the general public in detecting media risk as it only runs on asynchronous/offline videos and not in real-time/online environments, it is considered by this study to be successful in determining seizure-risk of videos; therefore, this study will use PEAT to determine ground truth values of whether produced media for training and testing of models contains seizure-triggers.

Previous studies have focused on analyzing fixed clips with epileptogenic media, which, while helpful in offline environments, fail to scale in real-time environments given the required computational intensity in the used methods \cite{b4, b6, b7}. Some studies also fail to preserve the quality or detail of the original content as a result of completely reconstructing the original media source to sanitize the video, most notably done in Barbu et al. with his use of Generative Adversarial Networks (GANs) to asynchronously, although successfully, reconstruct seizure-inducing media to 'sanitize' them from flashes \cite{b7}. 

Given that the current norm for preventing epileptic flashes in media is to detect asynchronously when flashes will occur in a video, then notifying the user, there is a lack of a real-time solution for dealing with seizure-causing content. Furthermore, current methods rely on 'prevention', rather than 'mitigation', which completely restricts the user from seeing the content, rather than obscuring only the dangerous elements of the content  \cite{b4, b8}. This study implements two approaches in a unified software system: first, a novel flash-trigger detection algorithm utilizing color differentials, and second, a mitigation system that sanitizes the media by preserving the underlying content while removing epileptogenic elements.

Crucially, this study provides novel insight into how intrinsic color properties are related to seizure risk and mitigation thresholds, calling for the adoption of broadened WCAG guidelines to better account for seizure risk.

\section{Methods}

\subsection{System Overview} 

To address the risk of seizures caused by flashing visual stimuli in modern media, we propose a two-part system. The first part detects seizure-inducing flashing in real time, while the second applies targeted mitigation filters to overlay the video, in order to reduce that flashing without altering the underlying detail of the media. The detection system is built on modeling the rate of change in color, and the mitigation system works by reducing luminance and smoothing color transitions.

\subsection{Flash Detection Theory}
In theory, in the event that the average rate of change of color, or average amount of flashing, in an area is sufficiently high enough, the detection model should evaluate a positive result in the respective area, meaning that the given area has seizure-inducing media. This event can be modeled with the following function:
$(F= \frac{dc}{dt}) > T$, where $F$ represents flashing, $c$ is the color in an area at a given time, so $\frac{dc}{dt}$ represents the rate of change of color at a given time, or the amount of flash $F$.  $T$ is the threshold for which the detection model learns the threshold of flash safety.

As such, when the average rate of change of color $\frac{dc}{dt}_{avg}$ exceeds a learned threshold, then the model should implement mitigation of flashing in the area.

\subsection{Color Representation}
Given the two main types of flashing (high luminance flashes and high color contrast flashes), this research proposes utilizing the CIELAB color space (interchangeably called the Lab* color space) for evaluating differences in color. Lab* is a perceptually uniform color space, used to better quantify differences in how humans perceive different colors \cite{b9}, as opposed to a perceptually non-uniform color space (for example, RGB). Unlike RGB, Lab* better aligns with how humans perceive differences in light and color. 

This study utilizes Lab* for the purposes of specifically seizure-risk detection and mitigation for two primary reasons:

\begin{enumerate}
    \item Computer-calculated differences in how humans perceive color will theoretically be more accurate when represented in perceptually uniform color spaces. This is to algorithmically mimic and quantify how PSE-individuals actually experience rapid flashes in color.
    \item Lab* has three axes determining color representation (L*, a*, and b*). Along the L* axis, the luminance (brightness) of colors differs, while along the a* and b* axes, color hue differs. Given this, the mitigation algorithm can isolate axes of color representation in flashing in order to independently mitigate components for flashing in each axis independently. 
\end{enumerate}

Representation of the event that amount rate of flashing $F$ exceeds the safe threshold $T$ can thus be rewritten as the following:
 $(F = \frac{dL}{dt} + \sqrt{{(\frac{da}{dt})}^2 + {(\frac{db}{dt})}^2}) > T$

Here, we can implement independent algorithms to mitigate the rate of change of luminance $\frac{dL}{dt}$ and the rate of change of color hue $\sqrt{{(\frac{da}{dt})}^2 + {(\frac{db}{dt})}^2}$.

\subsection{Trigger Detection System Method}
Given that no publicly available repository of seizure-causing videos is available to train the trigger detection model, a dataset of 1000 synthetic, ten-second videos, each with controlled variations, was generated. Variations included:
\begin{itemize}
    \item Whether the background was flashing
    \item At what rate did the background flash (if it was flashing)
    \item Whether there was a shape object in the foreground
    \item The size of the shape
    \item Whether the shape would flash in color
    \item At what rate the shape would flash at (if it was flashing)
\end{itemize}

To avoid inappropriate usage of the seizure-causing videos by the public, the dataset will not be publicly published. The dataset is available upon request to the author.

The training dataset was comprised of 800 of the 1000 videos. Each training video was analyzed using PEAT to determine if it posed a seizure risk. Using this dataset, we calculated the frame-to-frame average color change in CIELAB space to get the average amount of flashing $F_{avg}$. These rates were then used to train a logistic regression model, using PEAT’s binary risk output as the ground truth - if the video contained any frames with detected flashes, then it would be marked as "containing a flash-trigger". The logistic model learned the threshold value $T$ at which the amount of flashing $F$ should stay under, as videos with excessive flashing were likely to be marked by PEAT as containing seizure triggers. This produced a model which took the average rate of change of color, or amount of flashing, of the screen over time, and outputted whether this level of flashing was sufficient to pose a risk of seizure-inducement (binary output). 

\begin{figure}[htbp]
\centerline{\includegraphics[width=8cm]{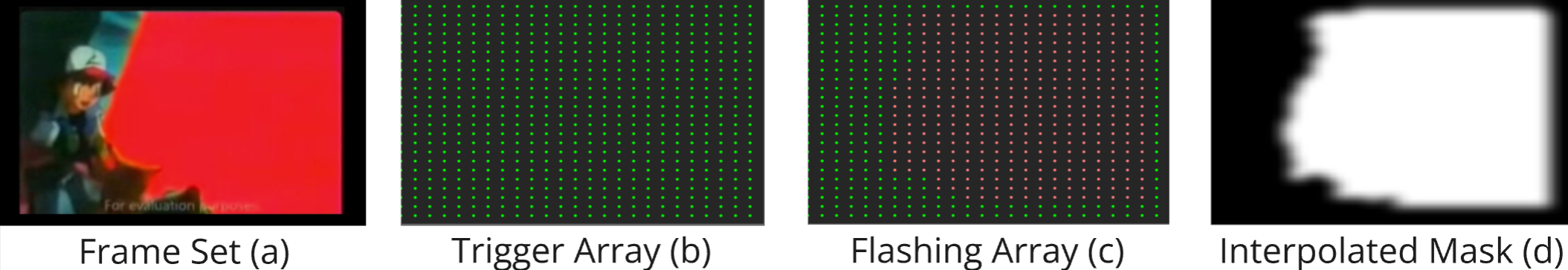}}
\caption{Visualization of the trigger array (b) applied and activated by famous seizure-causing Pokémon episode “Dennō Senshi Porygon” (1997), © The Pokémon Company (a), which caused hundreds of viewers to be hospitalized in Japan in 1997 \cite{b10}}
\label{fig-pokemontriggerarray}
\end{figure}

To account for the \textit{size} of flashing regions, we implemented a spatially distributed detection model across the screen, proposed by Alzubaidi et al. \cite{b6}. In this implementation, multiple instances of the logistic regression model run independently on streams of pixel values from uniformly distributed positions on the screen. These instances run in a 'pixel array' across the screen over time, each one independently triggering when a pixel stream crosses the level of safe flashing, allowing for the mapping of regions of seizure risk, rather than classifying the entire screen as a seizure risk (Fig. \ref{fig-pokemontriggerarray}). 

As an example, in Fig. \ref{fig-pokemontriggerarray}, the trigger array (b) has nodes activated in the flashing areas of the frames of the video (a), as illustrated in the activated flashing array (c). These activated nodes are interpolated to delimit the area (d) in which the mitigation algorithm should be applied.

In practice, this allows for computationally efficient detection of flashes across the screen. Additionally, areas not covered by the trigger array are too small to be significant in size, as detailed in PEAT guidelines, which only considers flashes greater than a quarter of a 341 x 256 rectangular area of the screen as dangerous \cite{b5}. In a trigger array that captures 50 pixels horizontally by 50 pixels vertically, significantly fewer pixels must be monitored compared to previous approaches, which would require all pixels in the screen to be monitored. In a 1024 by 768 pixel screen, using a 50 by 50 trigger array results in a 99.7\% reduction in computational intensity ($\frac{1024\times768 pixels-50\times50 pixels}{1024\times768 pixels}$), allowing for real-time detection.

\subsection{Mitigation Algorithm}

Once the model detects an unsafe flashing event in a region, the mitigation algorithm is implemented in that area in real-time. We design two separate mitigation strategies to address each type of flashing: one for luminance, one for color.

\subsubsection{Luminance Mitigation}

To suppress luminance-induced flashing, a darkening filter is applied over the region. The intensity of the filter is adaptive to the luminance of the color, which this study will later determine to be correlated to the underlying color.

\begin{figure}[htbp]
\centerline{\includegraphics[width=6cm]{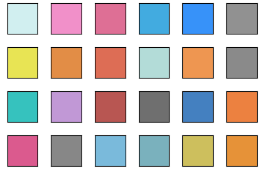}
}
\caption{Example of 24 of the 200 randomly chosen colors used as base color for white flash injection}
\label{fig200colors}
\end{figure}

We created 200 videos, each 10 seconds long and composed of a single, randomized base color (Fig. \ref{fig200colors}). White flashes were overlaid at a rate of 3 frames per second, with translucency of the white frames varying between separate tests from 10\% to 90\% in increments of 10\%.

\begin{figure}[htbp]
\centerline{\includegraphics[width=8cm]{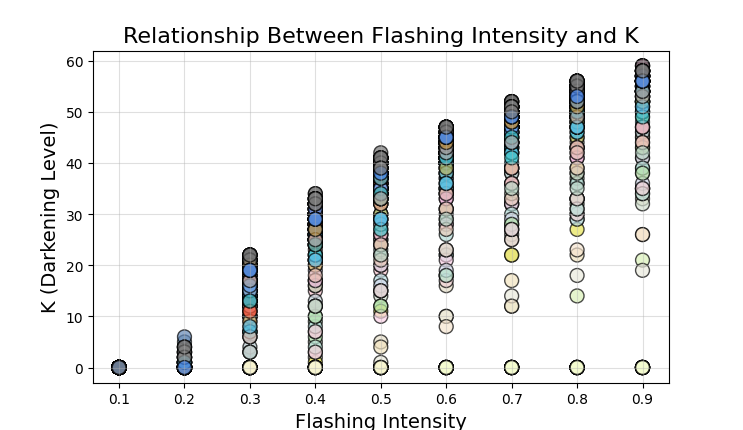}
}
\caption{Levels of flashing intensity across videos with different base colors versus the minimum required k-darkening level to mitigate flashing. Flashing intensity, or the percent opacity of the white flash injections, increases in increments of 10\%. The k-darkening level is equivalent to the percent opacity of darkness required to mitigate the injected flashes.}
\label{fig2}
\end{figure}

For each level of flash intensity, we applied a darkening filter at various levels of opacity, referred to as the “\textbf{k-level} of darkening”. The k-level directly equals the percent opacity of the black filter applied over the screen. For each level of flash intensity for each video, we determined the minimum k-level required to suppress flashing such that PEAT would mark the seizure-inducing video as non-seizure inducing. This allowed us to build a predictive correlative model of the minimum necessary darkening based on the base color value, given different flashing intensities (Fig. \ref{fig2}).

In real-time environments, exact flash intensity isn’t quantifiable. For this reason, we assume high-risk flashes occur around 70\% luminance and darken accordingly to balance safety and content visibility (preventing total darkening of the screen).

\subsubsection{Color Mitigation}

To reduce risk from high-color-contrast flashing, we applied a temporal color smoothing technique. This involved taking the running average color over the past \textbf{n }frames and applying that color as a semi-translucent overlay to the current frame, where \textbf{n} was arbitrarily chosen. In implementation, $\textbf{n}=15$ was chosen, meaning the average colors in the last 15 frames were averaged to one color, which was translucently applied to the area of flashing.

This method smooths out sudden color jumps while preserving underlying scene details. It also adapts automatically to the content, allowing for real-time color mitigation in the specific interpolated mask area (Fig. \ref{fig-pokemontriggerarray}(d)).

\section{Results}

\subsection{Detection Model Performance}
\begin{figure}
\begin{center}    
\begin{tikzpicture}[
box/.style={draw,rectangle,minimum size=2cm,text width=1.5cm,align=left}]
\matrix (conmat) [row sep=.1cm,column sep=.1cm] {
\node (tpos) [box,
    label=left:{Safe} ,
    label=above:{Safe},
    ] {74 videos};
&
\node (fneg) [box,
    label=above:{Risky}] {26 videos};
\\
\node (fpos) [box,
    label=left: {Risky}
] {14 videos};
&
\node (tneg) [box,] {86 videos};
\\
};
 \node [rotate=90,left=.15cm of conmat,anchor=center,text width=3.0cm,align=center] {\textbf{Actual Trigger Risk (PEAT)}};
\node [above=.05cm of conmat] {\textbf{Prediction outcome of detection model}};
\end{tikzpicture}
\caption{Correlation matrix of 200 test videos analyzed by the detection model. Risky videos were those detected to have a seizure-inducing flash by PEAT analysis.}
\label{fig3}
\end{center}
\end{figure}

The logistic regression model trained to detect seizure-inducing flashing achieved an overall classification accuracy of 80\% (160 out of 200 video classifications were correct) on the test set, which consisted of 200 10-second videos with the same randomized parameters as the original 1000 training videos, with 100 of the videos containing a potential seizure-trigger as determined by PEAT, and the other 100 videos lacking a seizure-trigger. The correlation matrix (Fig. \ref{fig3}) shows that the true negative classification rate, or the successful classification rate of videos without a flash trigger, was 74\% (74 out of the total 100 non-triggering videos were successfully ignored). Conversely, the true positive classification rate was 86\% (86 out of the total 100 videos containing a flash trigger were successfully detected as potentially dangerous), demonstrating high efficacy in practical, real-time environments. Additionally, a lower proportion of false classifications of risky media as safe (14\%) signals that the model successfully detects the majority of dangerous media.

The detection model also achieved an AUC-ROC score of 0.9038, indicating strong overall performance. The AUC (Area Under the Receiver Operating Characteristic Curve) quantifies the model's ability to discriminate between the classes, given the single parameter of averaged flashing in the frames over time. A value of 1.0 represents perfect classification, and 0.5 represents random guessing. A value above 0.9 is considered excellent, suggesting that the model is highly capable of distinguishing between triggering and non-triggering content using the quantification of flashing $F$ in the CIELAB color space.

To statistically confirm the effectiveness of the model, a proportion z-test was conducted:

\begin{itemize}
\item Proportion$_{model}$ = the proportion of correct classification of PSE-triggering videos
\item Null Hypothesis: p$_{model}=0.50$ (no better than random classification)
    \item Alternative Hypothesis: p$_{model}>0.50$ (classification was non-random)
\end{itemize}

When calculated with 80\% classification accuracy, $z=10.607$.
This corresponds to a $p-value < 0.001$, allowing us to reject the null hypothesis and conclude that the model performs significantly better than chance at detecting PSE-inducing content.

\subsection{Luminance Mitigation Results}

Luminance mitigation was found to be the most critical factor in reducing seizure risk, as measured by PEAT. When comparing the correlation between required \textbf{k-level} of darkening across CIELAB dimensions, the required k-level showed a strong inverse correlation with the color’s \textbf{L*} value (luminance) (Fig. \ref{fig5}). Higher luminance values caused decreased darkening applied, while lower intrinsic luminances of color (darker colors) resulted in more darkening being applied. Differences in color hues in A* and B*, although showing correlation to more darkening applied (Fig. \ref{fig2}), were not significantly learned by the model to affect darkening required (Fig. \ref{fig5}).

\begin{figure}
    \centering
    \includegraphics[width=0.9\linewidth]{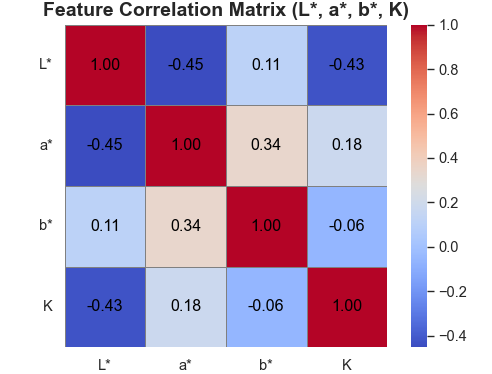}
    \caption{Feature Correlation Matrix for L*, a*, b* values of base colors across the 200 white flash-injected videos with the required k-level of darkening to mitigate flashing}
    \label{fig-labk-correlation}
\end{figure}

Specifically, the Pearson correlation coefficient between k and L* was -0.43, signifying that darker colors (lower luminance) required more darkening to achieve safety, likely due to their stronger contrast with the injected flashing white overlays. 
\begin{figure}
    \centering
    \includegraphics[width=1.0\linewidth]{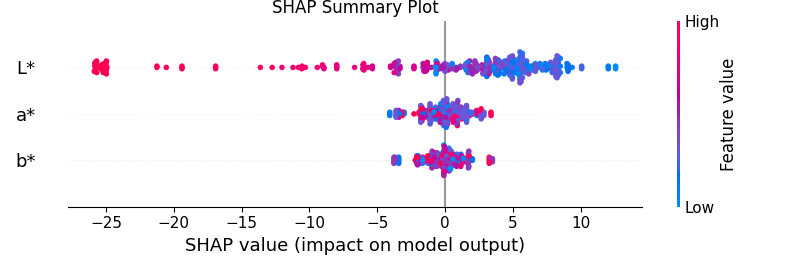}
    \caption{Shapley Additive explanations (SHAP) plot correlating how different CIELAB dimensions directly impact the model-predicted K-level of darkening.}
    \label{fig5}
\end{figure}

Additionally, intrinsically darker colors (blue, gray) required higher k-levels of darkening at higher flashing intensities in order to mitigate flashing, in comparison to intrinsically lighter colors (yellow, pink, light-green), which didn’t require increases in darkening in order to mitigate for higher levels of flashing intensity. This is likely because the intrinsically lighter colors, when compared to the white injected overlays, had little contrast with each other, and therefore, posed less of a flashing risk.

This is further supported by the non-negligible values in the Pearson correlation coefficient between L* and A*. This phenomenon is likely due to the Helmholtz-Kohlrausch effect, in which certain colors are perceived to be intrinsically brighter than other colors at the same actual brightness value \cite{b11}. This results in colors with different hues having different perceptual differences in brightness, as supported by consequently non-negligible Pearson coefficients between k-level and A* (Fig. \ref{fig-labk-correlation}), as darkening adjusts due to color hue (also visible in Fig. \ref{fig2} as darker colors require more darkening/higher k-level to mitigate flashing).

Interestingly, the correlation between L* and B* was negligible, even though the B* color hues scale from yellow to blue, which have more perceptual differences from each other than red to green (the scale for A* hues). However, this is likely because higher luminance values likely didn't correlate with many changes in perceptual difference - a darker yellow is relatively similar to a lighter yellow, as compared to a darker green with a lighter green. This also explains the lack of correlation with k-level and B*, as the lack of perceptual flash contrasts between frames required less darkening.

Conversely, given that blue of any brightness level contrasts greatly with white injections, blue would largely be considered darker regardless of starting luminance. The lack of correlation to k-level is also explained by this - values of blue, regardless of starting luminance, would always largely contrast with the injected white flashes, always requiring more darkening.

\subsection{Color Mitigation Results}
In addition to luminance suppression, the color mitigation algorithm, based on temporal color averaging, successfully mitigated seizure-inducing frames by reducing rapid chromatic shifts.

When applied to 200 high-risk videos (as flagged by PEAT), the combined system (luminance and color mitigation) was able to mitigate 92\% of flashing across frames, as evaluated by PEAT after the videos were treated with the unified detection and mitigation systems. This was calculated by dividing the number of resultant frames with flashing, as detected by PEAT, after being treated with the mitigation algorithms, by the initial number of flash frames detected by PEAT before sanitation. Importantly, the algorithm preserved the underlying visual content, unlike methods that discard or reconstruct entire frames (like in GAN-based 'sanitization' \cite{b7}). This result highlights the practical viability of this approach for real-world, real-time deployment on media platforms. 

\section{Discussion}

\subsection{Reframing the Understanding of Color in Seizure Triggers}

This study presents a new paradigm for understanding and addressing seizure triggers in digital media, particularly with respect to flashing stimuli and color contrast. Through the use of perceptually aligned color space analysis with CIELAB, we demonstrate that not all seizure-inducing flashes are created equal. Rather, perceptual contrast and intrinsic luminance of colors play a role in seizure causation, as demonstrated by the relationship between required darkening (k-level) at different levels of flashing intensity for different colors.

This study is the first to draw correlative implications between the Helmholtz-Kohlrausch effect and variations in seizure-causing color patterns. Intrinsic brightness in color should be considered when mitigating for seizure triggers in the future.

Additionally, due to this study's novel analysis done in the L*a*b* color space, independent correlations to luminance and color can be drawn, allowing for a deeper understanding of how intrinsic differences in color contribute to perceptual differences in seizure-risk flashes for photosensitive individuals. 

\subsection{Broadening Content Accessibility Guidelines}

Traditional research and policy standards, including the Web Content Accessibility Guidelines (WCAG), focus narrowly on limiting flash frequency and simple brightness thresholds, or specifically, red-saturated flashes. These standards stem from older understandings of PSE, which emphasized bright light flashes and red-dominant color events. However, the results of this study challenge that framework. WCAG focuses on red-saturated flashes, and while this study agrees that red-specific flashes continue to pose danger to PSE-individuals, this study calls for an expansion of this standard to encapsulate guidelines against flashes of, broadly, high-contrast flashing with attention to perceptual qualities of the color in said flashes.

While the WCAG are still largely not followed or enforced across the Internet, we hope that the broadening of WCAG, specifically Success Criterion 2.3.1, will emphasize a broader interpretation of the relationship between color types and flash intensity in seizure inducement.

\subsection{Enabling Practical, Real-Time Prevention}

Beyond theoretical insights, this study introduces a practical, real-time system for detecting and mitigating seizure-inducing stimuli. While prior approaches, such as GAN-based reconstructions (e.g., Barbu et al.), rely on regenerating video content in ways that are computationally heavy and likely to distort media, our method instead operates locally and in real-time, by darkening or color-smoothing only the high-risk regions. This preserves the integrity of the original media while substantially reducing seizure risk in the most dangerous parts of the screen.

The \textbf{92\% mitigation success rate}, achieved without compromising the underlying video content, marks a step forward in feasible, deployable protection for PSE patients, especially on platforms where user-uploaded or real-time content cannot be manually reviewed. These insights and implementations pave the way for stronger protections for PSE individuals from dangerous triggers in digital media, both in policy, as well as in software.

\section*{Acknowledgment}
The author Ishan Pendyala received guidance from Dr. Todd Masel, MD, who gave significant insights into how photosensitive epilepsy affects afflicted people every day. His voluntary support helps to answer the continually existing problems posed by photosensitivity. No other support was provided by this source or other sources.

\vspace{12pt}
\color{red}


\begin{thebibliography}{00}
\bibitem{b1}Fisher RS, Acharya JN, Baumer FM, French JA, Parisi P, Solodar JH, et al. Visually sensitive seizures: An updated review by the Epilepsy Foundation. Epilepsia. 2022 Feb 7;63(4):739–68.

\bibitem{b2} Shedding Light on Photosensitivity [Internet]. Epilepsy Foundation. 2021. Available from: https://www.epilepsy.com/stories/shedding-light-photosensitivity-one-epilepsys-most-complex-conditions

 
\bibitem{b3} Understanding Success Criterion 2.3.1 Understanding WCAG 2.0 [Internet]. W3.org. 2025. Available from: https://www.w3.org/WAI/GL/WCAG20/WD-UNDERSTANDING-WCAG20/seizure-does-not-violate.html

 
\bibitem{b4} South L, Saffo D, Borkin MA. Detecting and Defending Against Seizure-Inducing GIFs in Social Media. Association for Computing Machinery [Internet]. 2021 May 6; Available from: https://dl.acm.org/doi/10.1145/3411764.3445510 

\bibitem{b5}University of Maryland College of Information. Photosensitive Epilepsy Analysis Tool (PEAT) [Internet]. Trace Research \& Development Center. Available from: https://trace.umd.edu/peat/ 

\bibitem{b6} Alzubaidi MA, Otoom M, Al-Tamimi AK. Parallel scheme for real-time detection of photosensitive seizures. Computers in Biology and Medicine. 2016 Mar;70:139–47. 

\bibitem{b7} Barbu A, Banda D, Katz B. Deep video-to-video transformations for accessibility with an application to photosensitivity. Pattern Recognition Letters. 2020 Sep;137:99–107. 

\bibitem{b8} Wakefield J. TikTok offers feature to avoid seizure triggers. British Broadcasting Corporation [Internet]. 2020 Nov 25; Available from: https://www.bbc.com/news/technology-55071345 

\bibitem{b9} Suzuki T, Ito C, Kitano K, Yamaguchi T. CIELAB Color Space as a Field for Tracking Color-Changing Chemical Reactions of Polymeric pH Indicators. ACS Omega. 2024 Aug 15;9(34). 

\bibitem{b10} Takahashi T. Action needed to prevent “Pokemon seizures.” Neurology Asia [Internet]. 1998 Jan 1;14028(4):8. Available from: https://www.neurology-asia.org/articles/19991\_001.pdf

\bibitem{b11} Nayatani Y. Simple estimation methods for the Helmholtz-Kohlrausch effect. Color Research \& Application. 1997 Dec;22(6):385–401. 
\end{thebibliography}
\end{document}